# Magnetism of Hexagonal $Mn_{1.5}X_{0.5}Sn$ (X = Cr, Mn, Fe, Co) Nanomaterials

R. Fuglsby[1,2], P. Kharel[1,2*], W. Zhang[2,3], S. Valloppilly[2], Y. Huh[1] and D. J. Sellmyer[2,3]

[1]Department of Physics, South Dakota State University, Brookings, SD
[2]Nebraska Center for Materials and Nanoscience, University of Nebraska, Lincoln, NE
[3]Department of Physics and Astronomy, University of Nebraska, Lincoln, NE

Abstract

$Mn_{1.5}X_{0.5}Sn$ (X = Cr, Mn, Fe, Co) nanomaterials in the hexagonal $Ni_2In$-type crystal structure have been prepared using arc-melting and melt spinning. All the rapidly quenched $Mn_{1.5}X_{0.5}Sn$ alloys show moderate saturation magnetizations with the highest value of 458 emu/cm$^3$ for $Mn_{1.5}Fe_{0.5}Sn$, but their Curie temperatures are less than 300 K. All samples except the Cr containing one show spin-glass-like behavior at low temperature. The magnetic anisotropy constants calculated from the high-field magnetization curves at 100 K are on the order of 1 Merg/cm$^3$. The vacuum annealing of the ribbons at 550 $^o$C significantly improved their magnetic properties with the Curie temperature increasing from 206 K to 273 K for $Mn_{1.5}Fe_{0.5}Sn$.

-----------------------------------------

*parashu.kharel@sdstate.edu

I.      Introduction

Mn-based Heusler materials in the non-cubic crystal structure have attracted much current attention because some of the materials are predicted to have high spin polarization at the Fermi level and show large magnetocrystalline anisotropy with a Curie temperature well above room temperature [1, 2]. Many of these materials have relatively low saturation magnetization due to a competing ferro-and antiferromagnetic interactions between transition-metal moments located at different lattice sites [3, 4]. An interesting feature in these materials is that their magnetic properties can be modified to fit a specific application by adjusting the concentration of moment bearing elements in the alloys [5, 6]. High-anisotropy materials with moderate magnetization and high Curie temperature have applications both in spintronic devices and high-density recording [7, 8]. For spintronic application, high-anisotropy materials with high spin polarization at the Fermi level are desired. A typical example of a non-cubic Heusler alloy that has been extensively investigated recently is the tetragonal $Mn_xGa$ ($1 \leq x \leq 3$) system. The magnetic and electronic structure properties of these alloys, especially those of the tetragonal $Mn_3Ga$, are consistent with the requirement of the spin-transfer-torque-based spintronic devices [7, 9-12]. These findings have stimulated further research on other non-cubic Mn-based compounds from the Heusler family to understand their structural, magnetic and electron transport properties. Our interest is to understand the structural and magnetic properties of a hexagonal half-Heusler-type compound $Mn_2Sn$ and to investigate how these properties change if a fraction of Mn atoms is replaced by other transition metal atoms including Cr, Fe and Co.

$Mn_2Sn$ has been predicted to exist both in the cubic and hexagonal crystal structures, but only the hexagonal $B8_2$ phase ($Ni_2In$-type structure) with space group p63/mmc ($a = 4.37$ Å, $c = 5.47$ Å) has been synthesized experimentally [13,14]. The cubic phase (bcc L21 phase) has been predicted to be a half-metallic fully compensated ferrimagnet [15]. In the hexagonal $Mn_2Sn$ lattice, the Wyckoff's 2a (0, 0, 0; 0, 0, 1/2) and 2d (1/3, 2/3, 3/4; 2/3, 1/3, 1/4) positions are occupied by Mn atoms and the 2c (1/3, 2/3, 1/4; 2/3, 1/3, 3/4) positions are randomly occupied by Mn and Sn atoms. The magnetic lattice can be divided into two sublattices; sublattice A consisting of 2a sites and sublattice B consisting of 2c and 2d sites. Since the interatomic distances between the Mn moments are $Mn_A$-$Mn_A$ = 2.76 Å, $Mn_A$-$Mn_B$ = 2.89 Å and $Mn_B$-$Mn_B$ = 3.75 Å, the $Mn_A$-$Mn_A$, $Mn_A$-$Mn_B$, $Mn_B$-$Mn_B$ exchange interactions are expected to be respectively negative, negative and positive [14]. We note that the Mn-Mn interaction typically is positive if the distance between Mn moments is greater than 2.90 Å [16].

## II.    Experimental Methods

The $Mn_{1.5}X_{0.5}Sn$ (x = Cr, Mn, Fe and Co) ribbons were prepared using arc-melting, melt-spinning, and annealing. The nanostructured ribbons were produced by rapidly quenching the induction melted $Mn_{1.5}X_{0.5}Sn$ onto the surface of a rotating copper wheel in a highly pure argon-filled chamber. The wheel was kept at the tangential speed of about 28 m/s for all samples. The effect of annealing on the structural and magnetic properties of the ribbons was investigated by heat treatment in a tubular vacuum furnace with a base pressure of about $10^{-7}$ Torr at a temperature of 200 $^oC$, 350 $^oC$, 450 $^oC$ and 550 $^oC$ for 2 hours. The elemental compositions were confirmed using energy dispersive x-ray spectroscopy (EDX), which showed that the compositions were close to the estimated values within an error of 1 at. %. The structural properties of the samples were studied by x-ray diffraction (XRD) using a Rigaku MiniFlex 600 with Cu source. The magnetic properties were investigated with a Quantum Design vibrating sample magnetometer in a Physical Properties Measurement System [PPMS]. Some of the high-temperature magnetic measurements were done in a VersaLab magnetometer. The Rietveld analysis of the x-ray diffraction patterns was performed using TOPAS software [11].

III.    Results and Discussion

Fig. 1(a) shows the x-ray diffraction (XRD) pattern of the rapidly-quenched $Mn_{1.5}X_{0.5}Sn$ (x = Cr, Mn, Fe and Co) ribbons recorded at room temperature. The XRD patterns are indexed with the standard pattern of a hexagonal $Mn_2Sn$ compound in the $Ni_2In$-type crystal structure, without including any elemental or alloy secondary phases. As shown in Fig. 1(b), the Reitveld refined lattice parameters for $Mn_2Sn$ are $a = 4.40$ Å and $c = 5.48$ Å, where both $a$ and $c$ show a systematic change with the atomic radii of X elements in $Mn_{1.5}X_{0.5}Sn$. The substitution of Fe or Co for Mn in the 2a sites of $Mn_2Sn$ is expected to produce a significant decrease in the lattice parameters because of the smaller atomic radii of Fe and Co as compared to that of Mn, which is consistent with our XRD data. However, the occupation of 2d sites is likely to produce a minimal effect. The vacuum-annealed samples also have a hexagonal crystal structure, but the XRD patterns of the low-temperature annealed (200 °C and 350 °C) samples contain a few superstructure diffraction peaks, which are absent in the patterns of the rapidly quenched and high-temperature (450 °C and 550 °C) annealed ribbons. These superstructure lines can be indexed with lattice parameters $a' = 3a$ and $c' = c$ as reported in ref. [14].

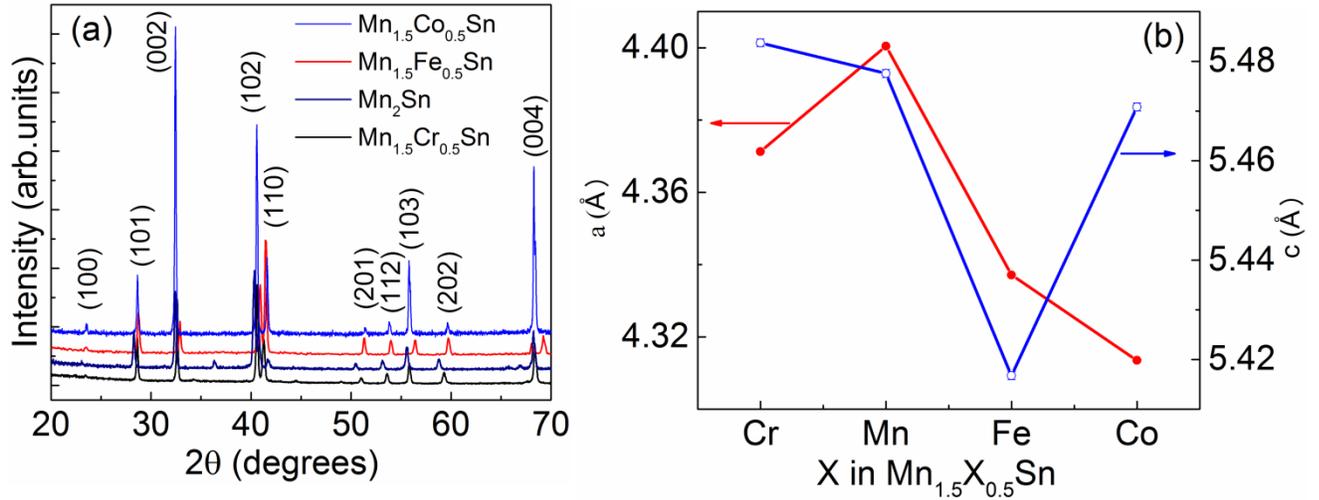

Fig.1: (a) Room-temperature x-ray diffraction patterns of rapidly quenched $Mn_{1.5}Cr_{0.5}Sn$, $Mn_{1.5}Mn_{0.5}Sn$, $Mn_{1.5}Fe_{0.5}Sn$ and $Mn_{1.5}Co_{0.5}Sn$ ribbons, respectively from bottom to top. (b) Lattice constants $a$ (scale on the left) and $c$ (scale on the right) as a function of X in $Mn_{1.5}X_{0.5}Sn$ (X = Cr, Mn, Fe, Co). The typical standard errors for the lattice parameters $a$ and $c$ are 0.0006 Å and 0.0008 Å, respectively. The diameters of the symbols in Fig. 1(b) are very close to the size of the error bars.

Fig. 2 shows the magnetic-field dependence of magnetization M(H) for the rapidly quenched $Mn_{1.5}X_{0.5}Sn$ (X = Cr, Mn, Fe, Co) ribbons measured at 100 K. The M(H) hysteresis loops of all $Mn_{1.5}X_{0.5}Sn$ ribbons except that of $Mn_{1.5}Co_{0.5}Sn$ are nearly saturated at 70 kOe, but the coercivities are relatively small ($H_c$ < 500 Oe). The magnetic properties of the $Mn_{1.5}X_{0.5}Sn$ alloys are highly sensitive to the type of X atom with the Fe-containing sample having the highest high-field (H = 70 kOe) magnetization of 458 emu/cm$^3$. The magnetizations at 30 kOe for all rapidly quenched and vacuum annealed samples are displayed in Table 1. These magnetic alloys are expected to show substantial magnetocrystalline anisotropy because of their non-cubic crystal structures. The magnetocrystalline anisotropy constants $K$ calculated from the high-field-magnetization curves measured at 100 K are on the order of 1 Merg/cm$^3$. As shown in the inset of Fig. 2, the anisotropy constant scales almost linearly with the number of valence electrons in the X atoms. The cobalt-containing sample shows the highest value of anisotropy constant K = 2.4 Mergs/cm$^3$. The anisotropy constant was calculated using the approach to saturation method [10, 17].

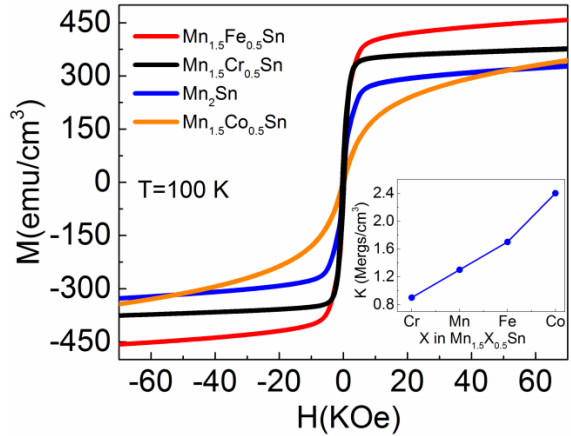

Fig. 2: Magnetic field dependence of magnetization of rapidly quenched $Mn_{1.5}X_{0.5}Sn$ (X = Cr, Mn, Fe, Co) ribbons measured at 100 K. The inset shows the anisotropy constant as a function of X in $Mn_{1.5}X_{0.5}Sn$ calculated from the high-field region of the M(H) curves recorded at 100 K.

The temperature dependences of magnetizations M(T) of the rapidly quenched $Mn_{1.5}X_{0.5}Sn$ alloys are shown in Fig. 3. The M(T) curves are similar to those of the ferro-or ferrimagnetic materials with a highest Curie temperature of about 206 K for the $Mn_{1.5}Fe_{0.5}Sn$ alloy. The Curie temperature was determined from the point where the extrapolated paramagnetic downturn in M(T) meets with the M = 0 line. As seen in Table 1, the Curie temperature in these samples roughly scales with the magnetization; the Fe-containing sample having the highest value of magnetization shows the largest value of the Curie temperature ($T_c$ = 206 K). A similar high value of Curie temperature, as compared to that of $Mn_2Sn$, has been observed in $Mn_{1.5}Cr_{0.5}Sn$ as well. This suggests that the presence of Fe or Cr in the $Mn_{1.5}X_{0.5}Sn$ lattice strengthens the positive exchange interaction. It is likely that the presence of Fe atoms in the 2a or 2d sites increases the number of ferromagnetically coupled Fe-Fe or Fe-Mn pairs breaking the antiferromagnetically coupled Mn-Mn chains. Similarly, the presence of Cr in the lattice sites may weaken the Mn-Mn antiferromagnetic interaction or may push some of the Mn atoms to the 2d sites resulting in the increase of magnetization. However, the relatively smaller magnetization and Curie temperature of the rapidly quenched $Mn_{1.5}Co_{0.5}Sn$ cannot be understood with this model. A remarkable feature in the M(T) curves of the $Mn_{1.5}X_{0.5}Sn$ alloys is that the M(T) curves of $Mn_2Sn$ and $Mn_{1.5}Fe_{0.5}Sn$ recorded at the FC (Field cooled) and ZFC (Zero-field cooled) conditions show a clear irreversibility below 18 K, indicating the presence of a spin-glass-like or spin-reorientation behavior. Interestingly, the spin-glass-like behavior has become more prominent in the Co containing sample with an increase in the ZFC/FC splitting temperature to 31 K but it is completely suppressed in the Cr containing sample. A possible explanation for this observation is that a competing ferro-and antiferromagnetic interactions between the magnetic moments at the 2a and 2d sites breaks the long-range spin chains leading to a spin-glass-like state at low temperature.

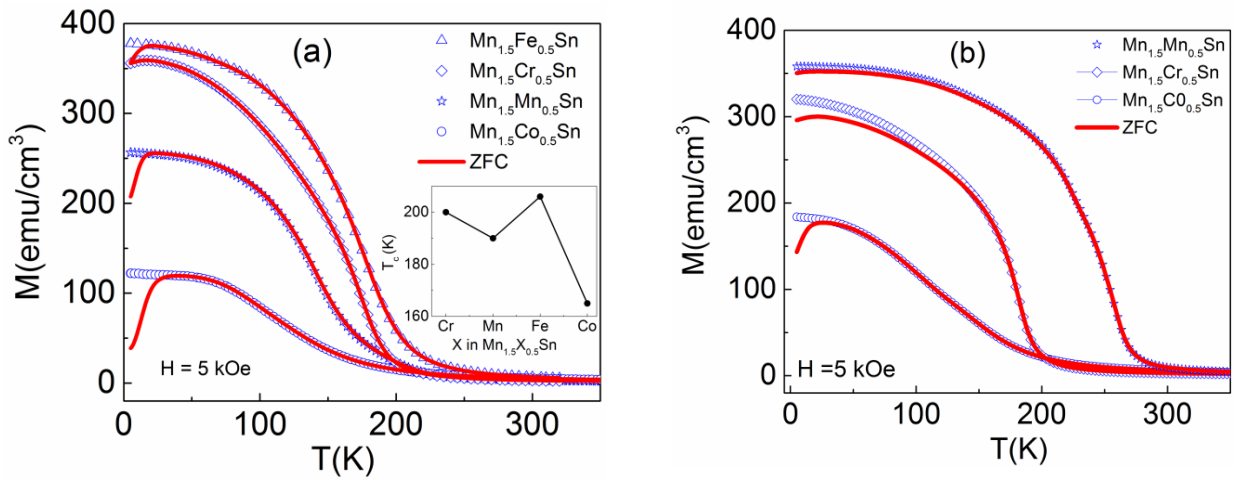

Fig. 3: (a) Temperature dependence of magnetization M(T) of rapidly quenched $Mn_{1.5}Co_{0.5}Sn$, $Mn_{1.5}Mn_{0.5}Sn$, $Mn_{1.5}Cr_{0.5}Sn$ and $Mn_{1.5}Fe_{0.5}Sn$ ribbons, respectively from bottom to top, measured at 5 kOe magnetic field. The inset plots the Curie temperature ($T_c$) as a function of X in $Mn_{1.5}X_{0.5}Sn$. The uncertainty in the values of $T_c$ is about 4 K. (b) ZFC/FC- M(T) curves of the vacuum annealed $Mn_{1.5}Co_{0.5}Sn$, $Mn_{1.5}Cr_{0.5}Sn$ and $Mn_{1.5}Mn_{0.5}Sn$ (from bottom to top) ribbons.

Table1: Magnetizations and Curie temperatures of $Mn_{1.5}X_{0.5}Sn$ alloys annealed at different temperatures. Average density used to express M in terms of $emu/cm^3 = 8.17$ $g/cm^3$

| Samples | $T_c$ (K) with samples annealed at | | | | | $M(emu/cm^3)$ at 30 kOe with samples annealed at | | | | |
|---|---|---|---|---|---|---|---|---|---|---|
| | None | 200°C | 350°C | 450°C | 550°C | None | 200°C | 350°C | 450°C | 550°C |
| $Mn_{1.5}Cr_{0.5}Sn$ | 200 | 200 | 194 | 187 | 188 | 368 | 310 | 245 | 245 | 302 |
| $Mn_{1.5}Mn_{0.5}Sn$ | 190 | 200 | 218 | 265 | 264 | 302 | 229 | 278 | 400 | 384 |
| $Mn_{1.5}Fe_{0.5}Sn$ | 206 | 197 | 231 | 263 | 273 | 433 | 392 | 384 | 392 | 400 |
| $Mn_{1.5}Co_{0.5}Sn$ | 165 | 172 | 190 | 179 | 194 | 270 | 245 | 245 | 253 | 245 |

We have investigated the effect of heat treatment on the magnetic properties of these alloys. The ribbons were annealed at a range of temperatures between 200 °C and 550 °C in a tubular vacuum furnace for 2 hours. The magnetic properties of the samples annealed at 450 °C and at 550 °C are very similar. The high-field (30 kOe) magnetizations measured at 100 K and the Curie temperatures of the alloys are displayed in Table 1. As shown in Table 1, there is a large increase in both the Curie temperature and magnetization of $Mn_2Sn$ alloy due to annealing. The magnetization and the Curie temperature of $Mn_2Sn$ have increased by about 100 $emu/cm^3$ and 74 K, respectively for the samples annealed at 450 °C. The Curie temperatures of $Mn_{1.5}Fe_{0.5}Sn$ and $Mn_{1.5}Co_{0.5}Sn$ alloys have also increased significantly due to annealing but without a noticeable change in the magnetization. The magnetization of $Mn_{1.5}Cr_{0.5}Sn$ has decreased significantly without much change in the Curie temperature. Further, the temperature of irreversibility between the ZFC and FC curves has increased to about 100 K in $Mn_2Sn$, $Mn_{1.5}Cr_{0.5}Sn$ and $Mn_{1.5}Fe_{0.5}Sn$ samples due to vacuum annealing, but the ZFC/FC splitting temperature of $Mn_{1.5}Co_{0.5}Sn$ ribbons is unchanged. The ZFC/FC-M(T) curves of vacuum annealed (at 450 °C) $Mn_{1.5}X_{0.5}Sn$ ribbons except that for X = Fe sample are shown in Fig. 3(b). Since the ZFC/FC curves below 100 K for $Mn_{1.5}Fe_{0.5}Sn$ and $Mn_2Sn$ are almost overlapping, the M(T) curve of Fe containing sample is not shown here to maintain the clarity of the figure. The large increase in magnetization and Curie temperature of $Mn_2Sn$ suggests that there is some rearrangement of Mn atoms in the $Mn_2Sn$ lattice due to vacuum annealing. Since there is ferromagnetic coupling between Mn atoms in the sublattice B, we assume that some of the Mn atoms from sublattice A may diffuse to sublattice B due to annealing. An interesting feature in the M(T) curves of the $Mn_2Sn$ ribbons is that the splitting between the ZFC and FC measurements is very small, although the splitting temperature has increased significantly. This may suggest that the origin of the FC/ZFC splitting in this system is caused by the spin disorder and/or vacancies in sublattice B.

IV.     Conclusions

$Mn_{1.5}X_{0.5}Sn$ (X = Cr, Mn, Fe, Co) nanomaterials in the hexagonal crystal structure were prepared using arc melting and rapid quenching in a melt spinner. The rapidly quenched $Mn_{1.5}X_{0.5}Sn$ alloys are ferro- or ferrimagnetic and the Curie temperature lies between 165 K for the Co-containing sample and 206 K for the Fe-containing sample. All the alloys show moderate saturation magnetization with a highest value of 458 emu/cm$^3$ for $Mn_{1.5}Fe_{0.5}Sn$. All the samples except Cr-containing sample show spin-glass-like behavior at low temperature. The vacuum annealing of the ribbons at 550 $^o$C significantly improved their magnetic properties; for example, the Curie temperature increased to 273 K for $Mn_{1.5}Fe_{0.5}Sn$. The observed magnetic properties are explained as the consequences of competing ferro- and antiferromagnetic coupling in the $Mn_{1.5}X_{0.5}Sn$ lattice. These materials will not find applications at room temperature or above because of their relatively low Curie temperatures. Further investigation is needed to determine if high doping levels with Fe, for example, can increase $T_c$ above room temperature.

Acknowledgements

This research is supported by NSF-MRSEC Grant DMR-0820521 (PK, RF, SV) and DOE-BES-DMSE Grant DE-FG 02-04ER46152 (WZ, DJS).YH is supported by Department of Physics SDSU.

References

1. V. Alijani, O. Meshcheriakova, J. Winterlik, G. Kreiner, G. H. Fecher, and C. Felser, J. Appl. Phys. **113**, 063904 (2013).
2. G. Kreiner, A. Kalache, S. Hausdorf, V. Alijani, J. F. Qian, G. Shan, U. Burkhardt, S. Ouardi, and C. Felser, Z. Anorg. Allg. Chem. **640**, 738 (2014).
3. T. Graf, C. Felser, S. S. P. Parkin, Prog. Solid State Chem **39**, 1-50 (2011).
4. J. Winterlik, S. Chadov, A. Gupta, V. Alijani, T. Gasi, K. Filsinger, B. Balke, G. H. Fecher, C. A. Jenkins, F. Casper, J. Kübler, G. Liu, L. Gao, S.S.P. Parkin, and C. Felser, Adv. Mater. **24**, 6283 (2012).
5. A. Nelson, Y. Huh, P. Kharel, V. R. Shah, R. Skomski and D J Sellmyer, J. Appl. Phys. **115**, 17A923 (2014).
6. S. Chadov, J. Kiss and C. Felser Adv. Funct. Mater. **23**, 832 (2013).
7. H. Kurt, K. Rode, M. Venkatesan, P. Stamenov, and J. M. D. Coey, Phys. Rev. B **83**, 020405(R) (2011).
8. Z Bai, L. Shen, G. Han, and Y. P. Feng, SPIN **2**, 1230006 (1012).
9. L. Zhu, S. Nie, K. K. Meng, D. Pan, J. Zhao, and H. Zheng, Adv. Mater. **24**, 4547 (2012).
10. Y. Huh, P. Kharel, E. Krage, R. Skomski, J. E. Shield, and D. J. Sellmyer, IEEE Trans. Magn. **49**, 3277 (2013).
11. Y. Huh, P. Kharel, V. R. Shah, X. Z. Li, R. Skomski and D. J. Sellmyer, J. Appl. Phys. **114**, 013906 (2013).
12. S. Mizukami, F. Wu, A. Sakuma, J. Walowski, D. Watanabe, T. Kubota, X. Zhang, H. Naganuma, M. Oogane, Y. Ando, and T. Miyazaki, Phys. Rev. Lett. **106**, 117201 (2011).
13. H. Shiraishi, T. Hori, N. Ohkubo, and K. Ohoyama, Phys. Stat. Sol.(c) **1**, 3660 (2004).
14. K. Yasukōchi, K. Kanematsu, and T. Ohoyama, J. Phys. Soc. Jpn. **16**, 1123 (1961).
15. H. Z. Luo, G. D. Liu, F. B. Meng, W. H. Wang, G. H. Wu, X. X. Zhu, and C. B. Jiang, Physica B **406**, 4245 (2011).
16. J M D Coey, J. Phys.: Condens. Matter **26**, 064211 (2014).
17. G. Hadjipanayis and D. J. Sellmyer, Phys. Rev. B **23**, 3349 (1981).